\documentstyle[aps,prl]{revtex}
\twocolumn

\draft
\begin{document}
\title{Scaling in Interaction-Assisted Coherent Transport}
\author{K. Frahm, A. M\"uller-Groeling, J.-L. Pichard and D. Weinmann}
\address{CEA, Service de Physique de l'Etat Condens\'e,
Centre d'\'Etudes de Saclay,\\
91191 Gif-sur-Yvette Cedex, France}
\maketitle

\begin{abstract}

The pair localization length $L_2$ of two interacting electrons
in one--dimensional disordered systems is studied numerically.
Using two direct approaches, we find
$L_2 \propto L_1^{\alpha}$,
where $L_1$ is the one-electron localization length and $\alpha \approx
1.65$. This demonstrates
the enhancement effect proposed by Shepelyansky, but
the value of $\alpha$ differs from previous estimates
($\alpha=2$) in the disorder range considered.
We explain this discrepancy using a scaling
picture recently introduced by Imry  and taking into account a
more accurate distribution than previously assumed for the
overlap of one-electron wavefunctions.

\end{abstract}
\narrowtext

Very recently, Shepelyansky \cite{shepel} considered the problem of
two interacting electrons in a random potential, defined by the
Schr\"odinger equation
\begin{eqnarray}
(E-V_{n_1, n_2})\psi_{n_1, n_2} &=&
\phantom{+} \psi_{n_1+1, n_2}+\psi_{n_1-1, n_2} \nonumber \\
& & +\psi_{n_1, n_2+1}+\psi_{n_1, n_2-1}.
\label{eq:SCHRO}
\end{eqnarray}

Here, $V_{n_1, n_2} = V_{n_1}+V_{n_2}+U \delta_{n_1, n_2}$,
$U$ characterizes the
on-site interaction and $V_i$ is a random
potential distributed uniformly in the interval $[-W,W]$.
The indices $n_1$ and $n_2$ denote the positions of the first and the
second electron, respectively. Shepelyansky proposed that,
as a consequence of the interaction $U$, certain eigenstates extend
over a range $L_2$ much larger than the
one--particle localization length $L_1 \propto W^{-2}$.
A key quantity in the derivation of
this spectacular result in one dimension is the matrix representation
${\cal U}$
of the Hubbard interaction in the disorder diagonal basis of localized
{\it one--electron} eigenstates. With $R_{n,i}$ the amplitude at site $n$ of
the one--particle eigenstate with energy $E_i$ we have
${\cal U}_{ij,lm}=U Q_{ij,lm}$
with
\begin{equation}
Q_{ij,lm} = \sum_{n}  R_{n,i} R_{n,j} R_{n,l} R_{n,m}.
\label{eqq}
\end{equation}
The $Q_{ij,lm}$ vanish unless all four eigenstates are roughly
localized within the same box of size
$L_1$. Assuming that $R_{n,i} \propto a_n/\sqrt{L_1}$ inside the box,
with $a_n$ a random number of order unity, and {\it neglecting} correlations
among the $a_n$ at different sites $n$, one finds
$Q \approx 1/L_1^{3/2}$  for a typical
nonvanishing
matrix element. In \cite{shepel}, this estimate was adopted and used
to reduce the original
problem to a certain band matrix model, eventually giving $L_2\propto L_1^2$.
Later, Imry \cite{imry} employed
the Thouless scaling block picture
to reinforce, interpret and generalize this result.
The key step in this approach involves the pair conductance
$g_2=(UQ/\delta_2)^2$, where $\delta_2$ is the two--particle level spacing in
a block of size $L_1$ and $Q$ is evaluated between adjacent blocks.
Using $Q \approx 1/L_1^{3/2}$ as before, Imry finds that $g_2 \approx 1$ on the
scale
$L_2\propto L_1^2$, in agreement with Shepelyansky. As a second important
result both approaches predict that the effect does not depend on the
{\it sign} of $U$.

In this letter, we confirm the enhancement effect by studying both
the original model (\ref{eq:SCHRO}) for finite size samples and
an infinite ``bag model'' with medium--range interaction. However,
we find $L_2\propto L_1^\alpha$ with $\alpha\approx 1.65$ instead of
$\alpha =2$ in both cases. Moreover, the sign of $U$ is not entirely
irrelevant. We suggest that the small value for $\alpha$
is due to a very peculiar
distribution of the $Q_{ij,lm}$. Imry's scaling approach, modified to
take into account the correct $Q$--distribution, is shown to be
more or less consistent with $\alpha \approx 1.65$ in one dimension.

{\it Transfer matrix approach for finite size chains.}
Eq.(\ref{eq:SCHRO}) is formally identical to a single--electron Anderson
model in two dimensions with symmetric (with respect to the diagonal
$n_1=n_2$) on-site disorder.
The indices $n_1$ and $n_2$ can be interpreted as cartesian
coordinates on a finite square lattice with $N^2$ sites.
The $N^2$ eigenstates of the Hamiltonian are either symmetric or antisymmetric
with respect to the interchange of $n_1$ and $n_2$. The interaction is
relevant for the symmetric states only (corresponding to electrons with
opposite spins). For the time being we disregard questions of symmetry
and proceed in analogy to the 2d Anderson model\cite{transfer,lyap}.
Imposing
hard--wall boundary conditions in the $n_2$--direction, transfer in the
$n_1$--direction is described by a product of $N$ matrices $M_p$.
This amounts to calculating the transport of one electron along the 1d
chain while the other one occupies some
state (or ``channel'') of the same chain. Note that only the energy $E$
of the {\it pair} of electrons is fixed. In numerical computations we are
restricted to finite sizes, $M_p$ being a $N\times N$ matrix. This limitation
is inessential
for $L_2<N$ and induces only minor finite--size corrections for the
transfer matrix eigenvalues otherwise \cite{lyap}.
We rewrite (\ref{eq:SCHRO}) in the form
\begin{equation}
\left[ \begin{array}{lr}
 u_{N+1}	\\
 u_N
\end{array} \right]= \prod_{p=1}^N M_p
\left[ \begin{array}{lr}
u_{1}	\\
u_0
\end{array} \right],
\end{equation}
where $u_p$ is a vector with N components $(\psi_{p, 1}, \ldots,
\psi_{p, N})$ and the transfer matrix $M_p$ is defined by:
\begin{equation}
M_p=
\left[ \begin{array}{cc}
(E-V_p) {\bf1} -  H_1	& \ - {\bf1}\\
{\bf 1} & \ {\bf 0}
\end{array} \right]
-U \left[ \begin{array}{cc}
\Delta_p	& \ {\bf 0}\\
{\bf 0} & \ {\bf 0}
\end{array} \right].
\label{Eq:MatM}
\end{equation}
Here, $H_1$ is the (transverse) single--particle Hamiltonian for the
second electron on the disordered chain of $N$ sites with hard--wall
boundary conditions, and $\Delta_p$ is a diagonal $N\times N$ matrix
with $(\Delta_p)_{ii} = \delta_{ip}$.

Before we numerically evaluate the set of decay lengths characterizing
this transfer matrix product, we insert a few qualitative remarks.
With $R$ the orthogonal $N\times N$ matrix diagonalizing $H_1$
(i.e., $H_1=RH_dR^T$, where $H_d$ is a diagonal matrix with entries
$E_\alpha$) we transform $M_p$ according to
\begin{equation}
{\cal M}_p=\left[ \begin{array}{cc}
R^T & \ {\bf 0}\\
{\bf 0} & \ R^T
\end{array} \right]
M_p
\left[ \begin{array}{cc}
R & \ {\bf 0}\\
{\bf 0} & \ R
\end{array} \right].
\end{equation}
In the noninteracting case ($U=0$)
${\cal M}(N)=\prod_{p=1}^N {\cal M}_p$ reduces
to $N$ decoupled products of $2\times2$ matrices and we deal
with a system of $N$ decoupled chains, all described by the
same Hamiltonian $H_1$. On each chain $\alpha$ the first
electron propagates with energy $E-E_\alpha$ while the second
one occupies an eigenstate of $H_1$ with energy $E_\alpha$.
The transfer matrix ${\cal M}_p$ defines $N$ localization
lengths $\xi_1(E-E_\alpha)$ related to $H_1$, eventually becoming
a continuous density in the thermodynamic limit. In the interacting
case ($U\ne 0$) we have to consider in addition the $N\times N$ matrix
$U  R^T \Delta_p R$ for each ${\cal M}_p$, which gives rise to two
modifications. First, this matrix couples the $N$ chains by a
hopping term
$U R^T_{\alpha, p}  R^{\phantom{T}}_{p, \beta}$
proportional to the
overlap (at site $p$) of two eigenstates of $H_1$ with energies
$E_\alpha$ and $E_\beta$, respectively. We recover the central
point of the present problem: the effect of the (local) interaction
is determined by the spatial overlap of the eigenstates of $H_1$.
If these states have random phases, the sign of $U$ will be
irrelevant in
the hopping terms. However, the second modification due to $U$,
a shift of $V_p$ on chain $\alpha$ by
$U  R^T_{\alpha, p}  R^{\phantom{T}}_{p, \alpha}$,
will remain sensitive
to this sign. In short, the interaction couples the 1d chains
and modifies (smoothens or enhances) the fluctuations of $V_p$ at
those sites where the eigenstates of $H_1$ are localized.

We have numerically studied $M(N)=\prod_{p=1}^N M_p$ as a function
of $U$ and $W$ at fixed $E$. We focus attention
on the logarithms $\gamma_i(N)=1/\xi_i(N)$ of the eigenvalues of
$M^T(N)M(N)$ divided by $2N$. One expects this set of inverse decay
lengths $1/\xi_i(N)$ to converge to a well--defined large--$N$ limit
\cite{lyap}. We denote by $L_2 \equiv \xi_1$ the largest of these lengths
to characterize the extension of the least localized quantum
state of the electron pair. In Fig.1 we present the five smallest and five
largest
inverse decay lengths $\gamma_i(N)$ for a given sample of size $N=100$
as a function of $U$.
The curves exhibit large sample--dependent fluctuations, but having
analyzed many different
samples we can draw the following conclusions: (i) the largest lengths
are increased by the interaction, confirming the effect discovered by
Shepelyansky; (ii) as expected for local interactions, shorter decay
lengths are less sensitive to $U$ since the electrons tend to be
localized with small or vanishing overlap; (iii) on average the curves
are symmetric around $U=0$ at the band centre $E=0$ (due to an exact
symmetry of the ensemble of Hamiltonians), but this symmetry is broken
for $E\ne 0$, and the sign of $U$ appears to be no longer irrelevant;
(iv) the nearest--neighbour spacing
distribution of the $1/\xi_i$ shows a crossover from Poisson statistics
(for $U=0$) to Wigner--Dyson statistics at finite $U$.

To determine the dependence of $L_2$ on $W$ we show in Fig.2 (note
the log scale) the ensemble--average
$ L_2(N) = 1/\langle \gamma_1(N)\rangle$
as a
function of $W$ for $N=100, E=0$, and $U=1$. One gets $L_2 \propto
W^{-2\alpha}$ with $2\alpha = 3.29$, a value which differs from the
estimate of Shepelyansky \cite{shepel} and Imry \cite{imry} ($\alpha=2$).
Although this qualitatively confirms the enhancement effect,
a certain (weak) sensitivity
of $L_2$ on the sign of $U$, and especially the value of $\alpha$, indicate
that the approximations employed in \cite{shepel,imry} were not accurate
enough. We note that for the smallest disorder values considered
($W=0.7,0.8$) $L_1$ is of the order of the system size.
Nevertheless a
strong enhancement effect prevails. This interesting observation supports
the claim \cite{pc} that persistent currents (in metallic systems)
are enhanced by interaction effects.

{\it Bag model with medium range interaction.}
To further substantiate our results, we turn our attention
to the ``bag model'' already mentioned in \cite{shepel}. In the
infinite square lattice in which each point $(n_1,n_2)$ corresponds to
a wave function $\psi_{n_1,n_2}$ of the two--electron system, we consider
now the transfer along the diagonal instead of the
$n_1$-direction. We associate with each center--of--mass coordinate
$n=n_1+n_2$ a ``tranverse'' vector $\phi_n$ containing $N_b$ two--particle
configurations with fixed $n$, but increasing distance $|n_1-n_2|$.
In contrast to the previous study (where we have ignored the symmetry
of the wave function) we assume
$\psi_{n_1,n_2}=\psi_{n_2,n_1}$ and consider only points in the upper
triangle of the scheme. The coupling between the $\phi_n$ can be written as
\begin{eqnarray}
E\phi_{2n} &=& h_{2n}\phi_{2n} + t^{(e)}
               \left( \phi_{2n+1}+\phi_{2n-1} \right), \nonumber \\
E\phi_{2n+1} &=& h_{2n+1}\phi_{2n+1} + t^{(o)}
               \left( \phi_{2n+2}+\phi_n \right).
\label{eq1}
\end{eqnarray}
The form of the matrices $h_i$ and $t^{(e,o)}$ follows directly from
(\ref{eq:SCHRO}) and we have to distinguish between $\phi_{2n}$
and $\phi_{2n+1}$ since only the former has an
entry on the diagonal $n_1=n_2$. The corresponding $2N_b\times 2N_b$
transfer matrices are given by
\begin{equation}
M_i^{(e,o)} =
\left(
\begin{array}{ccc}
t^{(e,o)-1} (E-h_i)  &     -1  \\
1                    &      0  \\
\end{array}
\right).
\end{equation}
They have to be iterated on a set of $N_b$ different start vectors
$\Omega^{(j)} = (\phi_u^{(j)},\phi_d^{(j)})$ which have to be orthonormalized
from time to time as the iteration progresses \cite{transfer}.
Summing up the logarithmic
increases of the norms of these vectors one calculates the set of $N_b$
Lyapunov exponents $\gamma_i=1/\xi_i$ of this transfer matrix. The start
vectors $\psi^{(i)}$ tend to the stationary distribution of the eigenvectors
of $M^TM$,
where $M=\prod_{i=1}^\infty M_i^{(e)}M_i^{(o)}$.

Unfortunately, the truncation of the transverse vectors $\phi_i$ in
the bag model introduces an artificial, long--range interaction between the
two electrons: They feel a hard wall whenever their relative separation
$\Delta n$ reaches its maximum value $(\Delta n)_{max}=2N_b$. A detailed
analysis of the bag model therefore meets with two problems: (i) to
determine the localization length $L_1$ for the noninteracting problem
and (ii) to attribute any observed enhancement of $L_2 = \xi_1 > L_1$
either to the (unphysical) bag interaction or to the (physical) Hubbard
interaction.

We have performed numerical calculations for $E=0$ and $N_b=100, 200, 300$ in
the disorder range $0.7 \le W \le 2.0$.
In our model the two electrons interact with $U=1$ whenever their distance
$\Delta n$ is smaller than a fixed value $(\Delta n)_{int}$. This modification
of the Hubbard interaction is designed
to enhance the physical effect
as compared to the unphysical bag interaction. We have chosen
$(\Delta n)_{int} = 20$ (to be compared with $(\Delta n)_{max} = 2N_b =
200,400,
600$). In all cases the full Lyapunov spectrum and the corresponding modes
(i.e. the average of the vectors $\phi_u^{2}$ over the full iteration process)
have been calculated up to a statistical accuracy of at least 5 percent.

Modes with dominant weight at
very small $\Delta n$ (maximal $\Delta n$) owe their existence to the
physical Hubbard interaction (the bag interaction).
Therefore Fig.3, where we show the modes corresponding to the two smallest
Lyapunov exponents $\gamma_1$ and $\gamma_2$,
demonstrates that the medium--range Hubbard--type interaction
is slightly more effective in assisting coherent
transport than the artificial bag interaction.
The dependence of $L_2$ on the disorder parameter $W$ is determined, as
for the previous case of finite systems, by plotting
$\log L_2=\log(1/\gamma_1)$ versus $\log W$.
The corresponding curve for $N_b=300$ is shown in Fig.2.
We find $L_2 \propto W^{-2\alpha}$ with $2\alpha = 3.42$.
This is in relatively good agreement with the previous model and confirms
that $L_2$ does not scale as $L_1^2$.

Finally, we compare in the inset of Fig.3 the smallest $36$ Lyapunov exponents
(for $N_b=100$)
with and without the Hubbard--type interaction. Two points deserve to
be mentioned:
(i) The interaction selectively reduces the smallest
Lyapunov exponents, demonstrating the profound effect
on the localization properties of the system.
(ii) The smallest exponents which are no longer appreciably
influenced by the interaction can serve (via $L=1/\gamma$) as a reasonably
good estimate for the localization length $L_1$ of the {\it noninteracting}
system. The corresponding modes are the first ones to propagate without
taking advantage of interaction effects.


{\it Q-statistics and scaling.}
To understand the discrepancy between our calculated and previously
estimated exponents, we return to the overlap matrix element (\ref{eqq}).
We  have calculated the $Q$--distribution (see Fig.4) for a 1d system
with $W=1.0$ and $N=200$ in the following way. Defining two adjacent boxes
of size $L_1 \approx 25/W^2 = 25$ in the middle of the system, we selected all
(noninteracting) eigenstates having their maximum in either of these
blocks. Then the distribution was determined by considering all $Q_{ij,lm}$
involving two wavefunctions from each block, respectively. We find deviations
from a Gaussian behavior in two respects. First, while the
off--diagonal terms $Q_{ij,lm}$ ($i\ne j$ and/or $l\ne m$) are distributed
symmetrically around zero, the diagonal elements $Q_{ii,jj}$ are always
positive. These latter contributions describe simultaneous transitions of
two electrons (with opposite spins) between state $i$ and state $j$.
Therefore, they involve only two wavefunctions (and not four), are typically
larger than the off--diagonal terms and preserve the sign of the interaction
$U$. Second, both the symmetric off--diagonal and the asymmetric diagonal
parts of the $Q$--distribution are very sharply peaked at zero and have
small but very long tails.
Fig. 4 suggests that the particular role of the diagonal
elements and the strongly non--Gaussian character of the distributions
might be responsible for the modified exponent $\alpha$. The form of $p(Q)$
affects both the arguments given by Shepelyansky and by Imry. The band matrix
model assumes that {\it all} pairs of two--particle states localized within
a distance $L_1$ are coupled with uncorrelated, {\it uniformly} distributed
interaction matrix elements. The scaling block picture estimates the
interblock coupling using essentially the same assumption. In reality,
sizable coupling matrix elements are {\it sparsely} distributed (and most
likely quite correlated) within $L_1$, so that the notion of a {\it typical}
interblock coupling
(or a well--defined bandwidth in the band matrix model)
is not unproblematic.

We nevertheless make an attempt
to improve previous estimates. Following Imry \cite{imry} we
consider two blocks of size $L_1$, each containing $L_1^2$ two--particle
states. We want to use the actual distribution $p(Q)$
and therefore have
to reduce a variable with a highly
non--trivial distribution to a ``typical'' value needed for the Thouless
scaling block picture. For illustration, we consider two extreme cases.
First we  calculate a {\it restricted} average $\langle Q^2 \rangle_{res}$
of the overlap matrix elements $Q$ coupling
a state in the first block and its nearest (in energy) neighbour in the
second block. Such an average is unaffected by the particular distribution of
the $Q_{ii,jj}$, which couple levels with spacings of the order of $1/L_1$
(instead of $1/L_1^{2}$), and turns out to be consistent with
$L_2\propto L_1^2$. Second, we calculate an {\it unrestricted} average
over the full spectrum, including also large energy separations. We
obtain  $\langle Q^2 \rangle \propto L_1^{-3.30}$ which eventually gives
$L_2 \propto L_1^{1.70}$ in good agreement with our independent
transfer matrix studies.
We believe that the second averaging procedure is more appropriate.
For a noninteracting system divided into blocks of size $L_1$ the Thouless
energy associated with a block is of the order of the one--particle
level spacing $\delta_1$.
The level uncertainty is  therefore large compared to the two--particle spacing
$\delta_2$ and should suffice to justify an unrestricted average
\cite{Altshuler}.


In conclusion, we have found that the pair localization length
$L_2$ scales with an exponent $\alpha\approx 1.65$, for a large
range of system parameters and in contrast to previous
predictions. The deviation from the expected value could be
traced back to the particular distribution of the overlap matrix
elements $Q$, composed of a symmetric off--diagonal and a positive diagonal
part, both sharply peaked with long tails.
By means of Imry's scaling picture the
anomalous exponent, characterizing the {\it interacting} electrons,
could be connected to properties of the {\it noninteracting} system.
The relevance of the $Q_{ii,jj}$ raises some doubts concerning
the independence of the effect on the sign of the interaction.
Studying the $Q$--distribution for $d>1$
might be a useful tool to investigate and
understand the effect discovered by Shepelyansky
also in higher--dimensional systems.

{\it Acknowledgments.}
We are indebted to Y. Imry for drawing our attention to the present problem.
We have also enjoyed fruitful discussions with F. v. Oppen, D. L. Shepelyansky,
and K. M. Slevin. This work was supported by fellowships of the DFG (K.F.),
NATO/DAAD (A.M.--G.) and the European HCM program (D.W.).

\begin{figure}


\caption{
The five smallest and five largest inverse decay lengths $\gamma_i(N)$
for a given sample with
$N=100$, $W=1$, and $E=2$ as a function of the Hubbard interaction $U$.
}
\label{fig1}
\end{figure}

\begin{figure}


\caption{
Log--log plot of the pair localization length $L_2$ as a function of the
disorder $W$ for (i) the finite chain with $N=100$
(solid) and (ii) the bag model with $N_b=300$ (dotted). Both curves
were calculated for $U=1$ and $E=0$. The exponent $2\alpha=3.3$
and the behavior of $L_1$ are also indicated.
}
\label{fig2}
\end{figure}

\begin{figure}


\caption{
The modes $\overline{\phi_u^2}$
corresponding to the two smallest Lyapunov exponents
$\gamma_1$ (solid) and $\gamma_2$ (dashed) for the bag model
with $N_b=200$, $W=1.5$, $U=1$, and $E=0$. Inset: The smallest
$36$ Lyapunov exponents for the same model, but $N_b=100$ and
(i) $U=1$ (solid) and (ii) $U=0$ (dotted).
}
\label{fig3}
\end{figure}

\begin{figure}


\caption{
The distribution of overlap matrix elements $Q$ for (i) a 1d chain with
$N=200$, $W=1.0$ (solid: diagonal terms, dashed: off-digonal terms) and
(ii) a normalized Gaussian (dotted) with the same variance as (i).
Inset: the same on a logarithmic scale.
}
\label{fig4}
\end{figure}

\end{document}